\documentclass[prb,11pt,notitlepage]{revtex4-2}

\usepackage{graphicx}
\usepackage{bm}
\usepackage{hyperref}
\usepackage{xcolor}
\usepackage{amsmath}
\usepackage{comment}
\usepackage{mathtools}

\def\be{\begin{equation}}
\def\ee{\end{equation}}
\def\ba{\begin{eqnarray}}
\def\ea{\end{eqnarray}}

\begin{document}

\title{Massive and Massless Quantum Cosmos}
\author{Rudolf Baier}
\email{baier@physik.uni-bielefeld.de}
\affiliation{Faculty of Physics, University of Bielefeld, D-33501 Bielefeld, Germany}
\author{Christian Peterson}
\email{cpeter16@uccs.edu}
\affiliation{Department of Physics, University of Colorado at Colorado Springs, Colorado Springs, CO 80918, USA}

\begin{abstract}
    The recent analysis of quantum cosmology by S. Gielen \cite{Gielen} is extended by discussing the case of dust (in the flat case). The dependence of the Wheeler-DeWitt equation on the operator ordering of the Hamiltonian in the case of a position dependent mass is explored, together with the $\Lambda$ dependence. As a main result, it is shown that matter enforces a quantized wave function as a solution of the corresponding Wheeler-DeWitt equation in the anti-de Sitter case.
\end{abstract}
\maketitle
\section{Introduction}

The interplay between quantum mechanics and general relativity stands as one of the most profound challenges in modern theoretical physics. Recently there has been an interest in applying quantum corrections to black hole and cosmological models \cite{Batic_2024,Gielen, Chowdhury, Vieira_2020, Ali}. This paper builds upon the framework of recent studies, such as those by S. Gielen \cite{Gielen}, to explore the dynamics of both massive and massless quantum cosmological models. Specifically, we investigate the effects of operator ordering in the Wheeler-DeWitt equation when the Hamiltonian includes a position-dependent mass. This approach enables us to examine the influence of matter density, represented by a dust-like component, and its impact on the wave function solutions in various cosmological scenarios, including those with different cosmological constants.

The physical Hamiltonian $H_\rho$ for the quantum cosmology, initially discussed by Vilenkin and others \cite{Kiefer2004-KIEQG,bojowald2011quantum,Vilenkin,kiefer2021impact}, is modified here to incorporate an energy density $\rho(a)$ (dust). We get
\be
    H_\rho = \frac{1}{2}\left[\frac{p_a^2}{a}+k a-\Lambda a^3 - \frac{8 \pi}{3}\rho(a)a^3 \right],
\ee
where the momentum is $p_a = a \dot{a}$. The FLRW equations $(G = c = 1)$ \cite{kiefer2021impact} are
\be
    \left(\frac{\dot{a}}{a}\right)^2 = \frac{8 \pi}{3}\rho(a)-\frac{k}{a^2}+\Lambda,
\ee
\be
    \ddot{a} = -\frac{4 \pi}{3}\rho(a) a +\Lambda a,
\ee
and
\be
   \dot{\rho}(a) = -3(\frac{\dot{a}}{a})\rho.
\ee
Note that $\frac{8 \pi}{3}\rho(a_0) = \rho_0$ and $\dot{a} = \frac{da}{dt}$. The Hamiltonian $H_\rho$, Eq.(1), is actually obtained by multiplying the FLRW Eq.(2), by a mass term $m/2$ \cite{rosen1993quantum}, but which has an $a$ dependence as $m = a$
\be
    \frac{m(a)}{2} \left[\dot{a}^2+k-\Lambda a^2 - \frac{8 \pi}{3} \rho(a) a^2\right].
\ee 
In the following analysis a flat universe is considered where $k = 0$. The constraint conditions $H_\rho = 0$ and $\dot{H_\rho}=0$ follow. Quantization of this model requires $p_a = -i\frac{\partial}{\partial a}$ where $\hbar = 1$, and imposing the Wheeler-DeWitt equation \cite{Kiefer2004-KIEQG,bojowald2011quantum,DeWitt}
\be
    H_\rho \psi(a) = 0.
\ee
To be explicit the operator ordering \cite{HAWKING1986185} is required for the kinetic part of the Hamiltonian \cite{Leblond}
\be
    \frac{1}{2}\frac{p_a^2}{a} = \frac{1}{2} a^\alpha p_a a^\beta p_a a^{\gamma = \alpha} \\
    = \frac{1}{2} p_a \frac{1}{a} p_a + \frac{1}{2}(\alpha + \gamma + \alpha \gamma) \frac{1}{a^3},
\ee
where $\alpha + \beta + \gamma = -1$ and
\be
    -\frac{1}{2} p_a \frac{1}{a} p_a = \frac{1}{2 a}(\frac{\partial^2}{\partial a^2} - \frac{1}{a}\frac{\partial}{\partial a}).
\ee
The Wheeler-DeWitt equation becomes
\be
\frac{1}{2}\left[\frac{d^2}{da^2} - \frac{1}{a}\frac{d}{da} - U(a)\right]\psi(a) = 0,
\ee
with
\be
    U(a) = (\alpha + \gamma + \alpha \gamma) \frac{1}{a^2} + a(-\Lambda a^3 - \rho_0).
\ee
A class of solutions of Eq.(9) depending on the operating ordering seen in Eq.(7) is discussed by Vieira et al \cite{Vieira_2020}. The structure of this paper is as follows: Section~\ref{sec:vilenkin ordering} reviews solutions to the Wheeler-DeWitt equation, E.(6), using Vilenkin ordering for both massless and massive cases. Section~\ref{sec:DeWitt} applies DeWitt ordering to Eq.(7), the kinetic part of the Hamiltonian, for a flat universe, examining the resulting wave function solutions. In Section Section~\ref{sec:Lambda}, we investigate the dependence of the Wheeler-DeWitt equation on the cosmological constant $\Lambda$, comparing solutions for positive, zero, and negative values. Finally, Section Section~\ref{ssec: Discussion} summarizes our findings and discusses their implications for the understanding of quantum cosmological models.
\section{Vilenkin ordering}
\label{sec:vilenkin ordering}
A simple form of the Wheeler-DeWitt equation is obtained by choosing $\alpha = \gamma = 0$, and $\beta = -1$ \cite{Vilenkin,Chowdhury}.
\subsection{massless case}
\label{ssec: Vil_A}
First we consider  the massless case where, $\rho_0 = 0$, the equation becomes \cite{gibbons1989typical}
\be
    \left[\frac{d^2}{da^2}-\frac{1}{a}\frac{d}{da}+\Lambda a^4\right]\psi(a) = 0.
\ee
Considering a negative $\Lambda$, we introduce a variable change as
\be
    z = \frac{1}{2}(-2\Lambda)^{\frac{1}{3}}a^2,
\ee
then the resulting equation becomes
\be
    \frac{d^2}{dz^2}\psi(z)-z\psi(z)=0.
\ee
Then the solution becomes, as derived by Vilenkin \cite{Vilenkin}, the Airy function
\be
    \psi(a) = Ai(z).
\ee

\subsection{massive case}
\label{ssec: Vil_B}
Next by including the mass term where, $\rho_0 \neq 0$, the equation reads
\be
    \left[\frac{d^2}{da^2}-\frac{1}{a}\frac{d}{da}+\Lambda a^4 + \rho_0 a\right]\psi(a) = 0.
\ee
We may introduce a variable change where
\be
    a = (\frac{3}{2}x)^{\frac{2}{3}},
\ee
and
\be
    \psi(a) = x^{\frac{1}{6}}\phi(x).
\ee
Substituting into Eq.(15) yields
\be
    \frac{d^2\psi}{dx^2}-\frac{1}{3x}\frac{d\psi}{dx}-\left(\frac{U(a)}{a}\right)_{a(x)}\psi(x)=0,
\ee
which simplifies to
\be
    -\frac{d^2\phi}{dx^2}+\frac{7\phi(x)}{36x^2}-\frac{9\Lambda}{4}x^2\phi(x) = \rho_0\phi(x).
\ee
Eq.(19) may be rearranged as an isotonic oscillator when keeping $\Lambda < 0$ and introducing a mass $m_0=1$, $g = \frac{7}{36 m_0}$, $\omega^2 = -\frac{9\Lambda}{4 m_0^2}$, so we have
\be
    -\frac{1}{2 m_0}\frac{d^2\phi(x)}{dx^2}+\frac{g}{2x^2}\phi(x)+\frac{m_0 \omega^2}{2}x^2\phi(x) = \frac{\rho_0 \phi(x)}{2 m_0}.
\ee
This is then reduced to
\be
    \phi''(x)+\left[\epsilon_n-\beta^2 x^2 -\frac{\alpha}{x^2}\right]\phi(x) = 0,
\ee
where, $\epsilon_n = \rho_0$, $\beta = m_0 \omega$, $\alpha = m_0 g > 0$  \cite{Ikhdair}.
One finds the energy eigenvalues of the isotonic oscillator as
\be
    E_n = (\rho_0)_n = \hbar \omega \left(2n+1+\frac{1}{2}\sqrt{1+\frac{4 m_0 g}{\hbar^2}}\right) = \hbar \omega (2n+1+\beta),
\ee
and the quantized wave function is given by associated Laguerre polynomials $L_n^\beta$,
\be
    \phi_n(x) \propto x^{\frac{1}{2}+\beta}\exp\left[-\frac{m_0\omega x^2}{\hbar^2}\right] L_n^\beta\left(\frac{m_0\omega}{\hbar^2}x^2\right),
\ee
where $x^2 \propto a^3$.
As a crosscheck one can obtain the massless case, $\rho_0 = 0$ from Eq.(19) to get
\be
    \Tilde{\phi''}(x)-\frac{7}{36x^2}\Tilde{\phi}(x)+\frac{9\Lambda}{4}x^2\Tilde{\phi}(x) = 0,
\ee
which has the solution,
\be
    \Tilde{\phi}\left(x = \frac{2}{3} a^{\frac{3}{2}} \right) = x^{-\frac{1}{6}} Ai \left( z = \frac{1}{2}(-2 \Lambda)^{\frac{1}{3}} a^2 \right).
\ee
This is consistant with section \ref{ssec: Vil_A}, and it can be seen that the solution has no eigenfunction property.
\section{DeWitt Ordering}
\label{sec:DeWitt}
\subsection{massive case}
In Eq.(6) the DeWitt ordering \cite{DeWitt} assumes that
\be
    \alpha = \gamma = -\frac{1}{4},\  \beta = -\frac{1}{2},\  \alpha + \gamma +\alpha \gamma = -\frac{7}{16},
\ee
and the Wheeler-DeWitt equation (flat case $k=0$) reads
\be
    \frac{1}{2}\left[\frac{d^2}{da^2}-\frac{1}{a}\frac{d}{da}-a\left(-\Lambda a^3 - \rho_0 - \frac{7}{16 a^2}\right) \right]\hat{\psi}(a) = 0.
\ee
As in section \ref{ssec: Vil_B} following the transformation in Eq.(14), Eq.(27) becomes
\be
    -\frac{d^2 \hat{\phi}(x)}{dx^2} + \frac{9 (-\Lambda)}{4}x^2 \hat{\phi}(x) = \rho_0 \hat{\phi}(x)
\ee
with $\omega^2 = 9 (-\Lambda)$ and $\rho_0 = E$ then the equation for the harmonic oscillator is obtained
\be
    -\frac{d^2 \hat{\phi}(x)}{dx^2} + \frac{\omega^2}{4}x^2\hat{\phi}(x) = E \hat{\phi}(x).
\ee
The solution ($\hbar = 1$) including the quantization condition is
\be
    E_n = (\rho_0)_n = \omega(n+\frac{1}{2}),\quad n = 0, 1, 2, ...
\ee
and the quantized wave function is given by Hermite polynomials,
\be
    \hat{\phi}(x) \propto H_n\left(X = \left(\frac{\omega}{2}\right)^{\frac{1}{2}}x\right).
\ee
Making the transformation $(-9 \Lambda)^{\frac{1}{4}} x = \hat{x}$ and $\alpha = \frac{\rho_0}{\omega}$ for Eq.(28) leads to
\be
    \frac{d^2\hat{\phi}(\hat{x})}{d\hat{x}^2} - \left(\frac{1}{4}\hat{x}^2-\alpha\right)\hat{\phi}(\hat{x}) = 0,
\ee
which is solved in terms of the Kummer confluent Hypergeometric functions \cite{Abramowitz} as
\be
    \hat{\phi}(\hat{x}) = \exp\left[\frac{-\hat{x}^2}{4}\right] {}_{1} F_1 \left(\frac{1}{4}-\frac{\alpha}{2}; \frac{1}{2}; \frac{1}{2}\hat{x}^2\right).
\ee
Taking $\frac{1}{4}-\frac{\alpha}{2} = -\frac{n}{2}$ then $(\rho_0)_n = \omega(n+\frac{1}{2})$ and the energy is seen to be quantized.
\subsection{massless case}
It is convenient to re-write Eq.(28) with $\rho_0 = 0$ and $(-9 \Lambda)^{\frac{1}{4}}x = \hat{x}$ to obtain
\be
    \left[\frac{d^2}{d\hat{x}^2} - \frac{1}{4}\hat{x}^2\right]\hat{\phi}(\hat{x})=0,
\ee
which is solved by the Hypergeometric function with
\be
    \hat{\phi}(\hat{x}) = \exp\left[\frac{-\hat{x}^2}{4}\right] {}_{1} F_1 \left(\frac{1}{4}; \frac{1}{2}; \frac{1}{2}\hat{x}^2\right),
\ee
which is not quantized \cite{Abramowitz}.
\section{\texorpdfstring{$\Lambda$}{Lg} Dependence}
\label{sec:Lambda}
So far only the anti-DeSitter case, $\Lambda<0$, has been considered in the previous sections. In the following, the Wheeler-DeWitt equation for $\Lambda \ge 0$ is considered by starting with Eq.(28)
\be
    \frac{d^2 \phi(x)}{dx^2}+\frac{9\Lambda}{4}x^2 \phi(x) + \rho_0\phi(x) = 0.
\ee
In the case of vanishing $\Lambda = 0$ the equation reduces to
\be
    \frac{d^2 \phi(x)}{dx^2}+\rho_0\phi(x) = 0.
\ee
This is solved by
\be
    \phi(x) = A e^{i \sqrt{\rho_0} x}+B e^{-i \sqrt{\rho_0} x},
\ee
which does not have any quantized energy levels. This indicates that the discrete energy levels for $\Lambda<0$ vanish with a vanishing $\Lambda$. For the DeSitter case, $\Lambda > 0$, one uses the transformation
\be
    x^2 = i (9 \Lambda)^{-\frac{1}{2}}\hat{x}^2,
\ee
to obtain the Kummer function (cf. Eq.(32))
\be
    \frac{d^2}{d\hat{x}^2}\phi(\hat{x})-\left(\frac{\hat{x}^2}{4}-i a_\rho \right)\phi(\hat{x})=0,
\ee
where $a_\rho = \frac{\rho_0}{3\sqrt{\Lambda}}$. The solution is in terms of Hypergeometric functions as
\be
    \phi(\hat{x}) = \exp\left[\frac{-\hat{x}^2}{4}\right] {}_{1} F_1 \left(\frac{1}{4}-i a_\rho; \frac{1}{2}; \frac{1}{2}\hat{x}^2\right),
\ee
which also has no real discrete energy levels.
\section{Discussion}
In summary, we investigated five cases of interest. We discovered that for $\Lambda < 0$, quantized solutions emerge in the massive scenarios, exhibiting oscillator-type behavior. Conversely, in the massless cases with $\Lambda < 0$, solutions do not exhibit eigenfunctions or eigenvalues. Free particle solutions arise when $\Lambda = 0$, while for $\Lambda > 0$, the solutions represent an inverted oscillator without quantized solutions as well. \cite{Ali}. 
\label{ssec: Discussion}
\appendix
\section{Path integral quantization}
In order to quantize the system with the Hamiltonian of Eq.(1) the path integral formalism is used \cite{Kaku:1993ym,Halliwell}. The path is given by a(t) and $p_a(t) = a \dot{a}$ and the action in terms of $H_\rho$ is given by
\be
    S(a,p_a) = \int dt \left[p_a(t) \dot{a}(t) - H_\rho(a,p_a)\right],
\ee
where
\be
    H_\rho(a,p_a) = \frac{p_a^2}{2a} + V(a).
\ee
The quantum mechanical partition function, where $\hbar = 1$ is given by
\be
    Z_{QM} = \int \frac{\mathcal{D} p_a}{2\pi}\int \mathcal{D}a \exp \left[i S(a,p_a)\right].
\ee
Performing the canonical transformation where $a = \left(\frac{3}{2}x\right )^{\frac{2}{3}}$ and $p_a = \left(\frac{3 x}{2}\right)^{\frac{1}{3}}p_x$ leads to $da dp_a = dx dp_x$ and $a\dot{a}^2 = \dot{x}^2$ to obtain
\be
    Z_{QM} = \int \frac{\mathcal{D} p_x}{2\pi}\int \mathcal{D}x \exp \left[i \int dt\left\{ p_x \dot{x} - \frac{p_x^2}{2}-V(x)\right\}\right].
\ee
Integrating over $p_x$ yields \cite{Kaku:1993ym}
\be
    Z_{QM} = \int \mathcal{D}x \exp \left[i \int dt \left\{\frac{\dot{x}^2}{2}-V(x)\right\}\right],
\ee
which is equal to
\be
    Z_{QM} = \int \mathcal{D}x \exp \left[i \int dt \ L(x,\dot{x})\right],
\ee
where L is the classical Lagrange function. This partition function leads to the equation
\be
    \left[-\frac{d^2}{dx^2}-\frac{9 \Lambda}{4}x^2\right]\hat{\phi}(x) = \rho_0 \hat{\phi}(x),
\ee
which is in agreement with Eq.(28), the case of the DeWitt operator ordering in section \ref{sec:DeWitt}. In this case there is no term proportional to $\frac{1}{x^2}$ that is present in contrast to Eq.(19) resulting from Vilenkin's ordering. The essence is the change of $S(a,p_a) \rightarrow S(x,p_x)$ which amounts to a specific operator ordering.
\section{Example for Curvature Index k = 1}
In order to simplify the operator ordering the problems seen in the kinetic terms of Eqs.(7-8) are 
 neglected\cite{Batic_2024,peleg1993quantum,Feinberg_1995},
\be
    \frac{p_a^2}{a} \rightarrow -\frac{1}{a}\frac{d^2}{da^2},
\ee
and taking $k = 1$ and $\Lambda = 0$ then the Wheeler-DeWitt equation reads,
\be
    -\frac{d^2\psi(a)}{da^2} + (k a^2 - \rho_0 a)\psi(a) = 0.
\ee
Transforming with
\be 
    x = a - \frac{\rho_0}{2},
\ee
gives
\be
    \left(-\frac{d^2}{dx^2}+x^2\right)\psi(x) = \frac{\rho_0^2}{4}\psi(x),
\ee
which has the quantized solution
\be
    \frac{\rho_0^2}{4} = (2n + 1), n = 0, 1, 2, ...
\ee
and the wave function in terms of Hermite polynomials \cite{Abramowitz},
\be
    \psi_n(x) \propto \exp\left[-\frac{x^2}{2}\right]H_n(x).
\ee
For the massless case, $\rho_0 = 0$, the solution becomes \cite{Abramowitz}
\be
    \psi(x) = \psi(a) \propto \exp\left[\frac{-a^2}{2}\right]  {}_{1} F_1 \left(\frac{1}{4}; \frac{1}{2}; a^2\right).
\ee
\begin{acknowledgements}
R.B. wishes to thank D.J. Schwarz for useful conversations. This work was supported by the Department of Energy, DOE award No DE-SC0017905. 
\end{acknowledgements}
\bibliography{References}
\end{document}